\documentclass[preprint,aps]{revtex4}

\usepackage{graphicx}

\begin{document}


\title{Coexistence of Two Sharp-Mode Couplings and Their Unusual Momentum Dependence in the Superconducting State of
Bi$_2$Sr$_2$CaCu$_2$O$_8$$_+$$_\delta$ Superconductor Revealed by Laser-Based Angle-Resolved Photoemission}
%
%
%
\author{Junfeng He$^{1}$, Wentao Zhang$^{1}$, Jin Mo Bok$^{2}$, Daixiang Mou$^{1}$, Lin Zhao$^{1}$, Yingying Peng$^{1}$, Shaolong He$^{1}$,   Guodong Liu$^{1}$, Xiaoli Dong$^{1}$, Jun Zhang$^{1}$, J. S. Wen$^{3}$, Z. J. Xu$^{3}$, G. D. Gu$^{3}$, Xiaoyang Wang$^{4}$, Qinjun Peng$^{4}$, Zhimin Wang$^{4}$, Shenjin Zhang$^{4}$, Feng Yang$^{4}$, Chuangtian Chen$^{4}$, Zuyan Xu$^{4}$, H.-Y. Choi$^{2}$, C. M. Varma$^{5}$ and X. J. Zhou$^{1,*}$
}

\affiliation{
\\$^{1}$National Laboratory for Superconductivity, Beijing National Laboratory for Condensed Matter Physics, Institute of Physics, Chinese Academy of Sciences, Beijing 100080, China
\\$^{2}$Department of Physics and Institute for Basic Science Research, SungKyunKwan University, Suwon 440-746, Korea
\\$^{3}$Condensed Matter Physics and Materials Science Department, Brookhaven National Laboratory, Upton, New York 11973, USA
\\$^{4}$Technical Institute of Physics and Chemistry, Chinese Academy of Sciences, Beijing 100080, China
\\$^{5}$Department of Physics and Astronomy, University of California, Riverside, California 92521
}
\date{October 2, 2012}

\begin{abstract}
Super-high resolution laser-based angle-resolved photoemission measurements have been carried out on Bi$_2$Sr$_2$CaCu$_2$O$_8$$_+$$_\delta$ (Bi2212) superconductors to investigate momentum dependence of electron coupling with collective excitations (modes). Two coexisting energy scales are clearly revealed over a large momentum space for the first time in the superconducting state of an overdoped Bi2212 superconductor. These two energy scales exhibit distinct momentum dependence: one keeps its energy near 78 meV over a large momentum space while the other changes its energy from $\sim$40 meV near the antinodal region to $\sim$70 meV near the nodal region. These observations provide a new picture on momentum evolution of electron-boson coupling in Bi2212 that electrons are coupled with two sharp modes simultaneously over a large momentum space in the superconducting states.  Their unusual momentum dependence poses a challenge to our current understanding of electron-mode-coupling and its role for high temperature superconductivity in cuprate superconductors.

\end{abstract}

\pacs{74.72.Gh,	74.25.Jb, 79.60.-i, 71.38.-k}

\maketitle

The physical properties of materials are dictated by the interaction of electrons with other entities like other electrons, phonons, magnons, impurities and etc. In particular, the superconductivity of materials involves interaction between electrons that may be direct or indirect through an exchange of a collective excitation (boson) that gives rise to electron pairing\cite{PWAnderson}. Understanding such many-body effects is essential for understanding the anomalous normal-state properties and superconductivity mechanism in high temperature cuprate superconductors. With a dramatic improvement of instrumental resolutions, angle-resolved photoemission spectroscopy (ARPES) has become a powerful tool to directly probe many-body effects in cuprate superconductors\cite{ZhouReview}. One prominent case is the observation of dispersion kink along the (0,0)-($\pi$,$\pi$) nodal direction. This nodal kink is found to be at $\sim$70 meV which is ubiquitous in different materials, different dopings and at temperatures both above and below the superconducting transition\cite{Bogdanovkink,Johnsonkink,Kaminskikink,Lanzarakink,Zhoukink,KordyukT}.  On the other hand, near the ($\pi$,0) antinodal region, a clear dispersion kink at $\sim$40 meV has also been identified\cite{GromkoANKink,KimANKink,SatoANKink,CukANKink}. In addition to the origin  of the nodal 70 meV kink and antinodal 40 meV kink that remain under debate (magnetic, phononic, or others), a long-standing puzzle to be resolved is their relationship. In particular, how does the 70 meV nodal kink evolve into the antinodal 40 meV kink when the momentum gradually moves from the nodal to the anti-nodal regions? Investigation of the momentum-dependence of the electron interaction is critical in understanding the electron pairing in high temperature superconductors\cite{Kamimura,Krahl,HCZ}.

In this paper we carried out super-high resolution ARPES measurements on Bi$_2$Sr$_2$CaCu$_2$O$_8$$_+$$_\delta$ (Bi2212) high temperature superconductors to study the momentum evolution of the electron interactions, with a particular aim to elucidate the relationship between the nodal 70 meV kink and the antinodal 40 meV kink. In an overdoped Bi2212 sample, we have revealed for the first time that two prominent energy scales coexist in a large area of momentum space in the superconducting state.  The $\sim$78 meV energy scale is observed over the Fermi surface from nodal to antinodal regions with nearly a fixed energy, while the other energy scale evolves from the antinodal region to the nodal region with its energy varying from $\sim$40 meV to $\sim$70 meV. These two energy scales also exhibit distinct temperature and doping dependence.  These observations provide a new picture on electron-boson coupling in Bi2212. It also provides insight on the unusual nature of these two electron-mode coupling and their possible role in high temperature superconductivity.

The angle-resolved photoemission measurements were carried out on our vacuum ultra-violet (VUV) laser-based angle-resolved photoemission system\cite{LiuIOP}. The photon energy of the laser is 6.994 eV with a bandwidth of 0.26 meV. The energy resolution of the electron energy analyzer (Scienta R4000) is set at 1 meV, giving rise to an overall energy resolution of 1.0 meV which is significantly improved from the previous ARPES measurements\cite{Bogdanovkink,Johnsonkink,Kaminskikink,Lanzarakink,Zhoukink,KordyukT,GromkoANKink,KimANKink,SatoANKink,CukANKink}. The angular resolution is $\sim$0.3$^\circ$, corresponding to a momentum resolution  $\sim$0.004 $\AA$$^{-1}$ at the photon energy of 6.994 eV, which is also improved from 0.009 $\AA$$^{-1}$ at a typical photon energy of 21.2 eV for the same angular resolution. High quality Bi2212 single crystal samples\cite{Supplementary}, including a slightly underdoped one with a T$_c$ at 89K (denoted hereafter as UD89K), an optimally-doped one with a T$_c$ at91 K (denoted as Opt91K) and an overdoped one with a T$_c$ at 82 K (denoted as OD82K),  were cleaved {\it in situ} and measured in vacuum with a base pressure better than 5$\times$10$^{-11}$ Torr.

Figure 1 shows the photoemission data on the Bi2212 OD82K sample measured along an off-nodal momentum cut (inset of Fig. 1a), at 17 K in the superconducting state. Due to much improved instrumental resolution and data quality, two coexisting energy scales can be identified for the first time from the second-derivative image (Fig. 1b), the effective electron self-energy (Fig. 1e) and the photoemission spectra (energy distribution curves, EDCs, Fig. 1f).  In the EDC second-derivative image (Fig. 1b),  there are two discontinuities appeared in the spectral weight distribution of the main band, accompanied by two spectral dips marked by arrows near two white lines (spectral peaks) on the right side.  In the corresponding effective real part of electron self-energy (Fig. 1e), two peaks emerge: one strong peak at $\sim$50 meV and another weak peak near 78 meV. Particularly, the photoemission spectra (EDCs, Fig. 1f) show clear multiple ``peak-dip-hump" structure with two dips that are revealed for the first time. For convenience, we call hereafter the energy scale at $\sim$50 meV as the LW energy scale, while the other one at $\sim$78 meV as the HI energy scale.

Detailed momentum dependent ARPES measurements have been carried out to investigate how the two coexisting energy scales evolve when the momentum moves from the nodal to anti-nodal regions. Fig. 2(a1-a5) shows the measured data along five momentum cuts; their corresponding EDC second-derivative data are shown in Fig. 2(b1-b5). Except for the nodal cut 1 where the two energy scales are too close to be distinguished (Fig. 2b1), the two coexisting energy scales are clear in all the other images, characterized by spectral weight discontinuities and two white lines that accompany the spectral dips (Fig. 2(b2-b5)). It is surprising to note that, for the cut 5, the two white lines appear even below the band bottom (Fig. 2b5). Two-dip structures are also clear in EDCs at the Fermi momenta for all the cuts except for the nodal one because the two energy scales become too close (Fig. 2c). The two energy scales can also be identified in the effective real part of electron self-energy (Fig. 2d) as indicated by the black arrow (for the HI energy scale) and red arrow (for the LW energy scale). We note that there are more fine features in the extracted electron self-energy (Fig. 2d), such as the 115 meV dip, 150 meV hump and others in the nodal cut as reported before\cite{Zhangfeatures}. However, in this paper, we will only concentrate on the two coexisting LW and HI energy scales that are revealed for the first time and their momentum evolution.  It is interesting that the LW and HI energy scales exhibit different momentum dependence, as shown in Fig. 2. Fig. 2e summarizes the position of these two energy scales determined from the peak position of the effective real part of electron self-energy (Re$\Sigma$) (Fig. 2d) and the dip position in the corresponding EDCs (Fig. 2c). The HI energy scale stays near 78 meV and varies little with momentum, while the LW energy scale varies obviously with momentum, dropping from $\sim$67 meV for the nodal cut to $\sim$40 meV (for $\Phi$$\sim$12 degree cut) near the anti-nodal region.

Temperature-dependent measurements provide further information on the distinct behaviors of the two energy scales, as shown in the effective real part of the electron self-energy measured at different temperatures for three typical momentum cuts (Fig. 3). It is clear that both the HI and LW energy scales exhibit dramatic superconductivity-induced electron self-energy change upon entering the superconducting state. However, they show different temperature-dependent behaviors in a couple of respects.  First, the coupling strength of the two energy scales exhibit strong momentum dependence, as evidenced by the temperature-induced self-energy change (Fig. 3(g-i)).  For the cut 3 near the antinodal region, the superconductivity-induced self-energy change is mainly dominated by the LW energy scale (Fig. 3i). When moving to the cut 2, the development of both energy scales is clear, but the self-energy change for the LW energy scale gets weaker. For the cut 1, in spite of a large superconductivity-induced self-energy change and apparent existence of the HI energy scale in the normal state, it is difficult to tell whether such a change is caused by HI or LW energy scale, or both because they are too close in energy to be distinguished. It appears that the coupling strength of the LW energy scale gets weaker when the momentum moves from the antinodal to the nodal regions.  Second, these two energy scales show different temperature dependence across T$_c$. It is clear from Fig. 3(g,h) that the HI energy scale is present even in the normal state above T$_c$. The temperature-induced self-energy enhancement starts to take off even above T$_c$ (82 K), such as the 90 K data in Fig. 3g. In contrast, no clear signature of the LW energy scale is observed in the normal state (Fig. 3i) and the obvious self-energy enhancement occurs well into the superconducting  state (like the data at 50 K and below).

It is important to check whether the coexisting two energy scales is a common phenomenon for the Bi2212 samples with different doping levels. Fig. 4(a-c) shows effective real part of electron self-energy along several momentum cuts of Bi2212 OD82K, Opt91K and UD89K samples, and the EDCs at k$_F$ along a typical cut ($\Phi$=18 degree) are shown in Fig. 4d. It is clear that the LW energy scale and its momentum dependence in the Opt91K (Fig. 4b) and UD89K (Fig. 4c) Bi2212 samples are similar to that in the OD82K sample (Fig. 4a). As for the effective electron self-energy, the signature of the HI energy scale is rather weak in the Opt91K and UD89K samples when compared with the OD82K case, but hint of its existence is still there. Particularly, one can still observe double-dip signature in the EDCs of the Opt91K and UD89K samples (Fig. 4d) even though it becomes less pronounced than that in the OD82K sample. Previous measurements on the optimally-doped T$_c$=91K sample also resulted in a similar temperature dependence of the effective electron self-energy along the nodal direction (Fig. 4 in \cite{Zhangfeatures}), as that found in the OD82K sample (Fig. 3g). These indicate that the two coexisting energy scales are common in the Bi2212 samples we have measured;  its stronger manifestation in the overdoped OD82K sample facilitates in identifying them in other samples. Overall, the LW energy scale for different Bi2212 sample follows a similar trend that it drops in its energy when moving from the nodal region to the antinodal region (Fig. 4e).

Since it is the first time to observe such two energy scales coexisting over a large momentum space in the superconducting state of Bi2212, it is important to understand their origin.  One possibility is the electron coupling with a sharp boson mode (with an energy $\Omega_0$)  plus a van Hove singularity near the anti-nodal ($\pi$,0) region at an energy position of E($\pi$,0). It was shown that two discontinuities may be generated in the electron self-energy at $\Omega_0$+E($\pi$,0) and $\Omega_0$+$\Delta_0$ (with $\Delta_0$ being a superconducting gap)\cite{Sandvik}. However, in this case, the resultant multiple energy scales are expected to be momentum-independent. In particular, the energy difference between the two scales, E($\pi$,0)-$\Delta_0$, is expected to be a constant so that the two energy scales should follow a similar momentum-dependence. These make one-mode scenario unlikely to account for our observation that the HI energy scale is basically momentum-independent in energy while the LW one exhibits clear momentum dependence in energy position (Fig. 4e). Distinct momentum and temperature dependence of the two energy scales point to their different origin.  The above behaviors indicate that, in the superconducting state, most likely electrons are coupled with two prominent sharp modes. This can be better illustrated by a simple simulation of the single particle spectral function for the electron interaction with two sharp bosonic modes. Fig. 1c shows the EDC second derivative image of the simulated one particle spectral function by considering electron interaction with two boson modes at 50 meV and 78 meV\cite{Supplementary}. Even with such a simple model simulation, the similarity of the measured (Fig. 1b) and simulated (Fig. 1c) images is obvious; the simulated image has captured the major features in the measured data.  With such a model simulation, it is clear that the position of the spectral dips in the EDC second-derivative (50 meV and 78 meV in Fig. 1b), the peak position in the effective real part of electron self-energy (50 meV and $\sim$78 meV in Fig. 1e) and the dip positions in the EDCs (50 meV and $\sim$78meV) represent the energy scales of the sharp modes involved; and indeed the energy scales extracted from these approaches are consistent with each other.

Our present work offers a new picture on the momentum evolution of the electron coupling in the Bi2212 superconductor. Previously, a single $\sim$70 meV energy scale observed near nodal region is assigned to either A$_{1g}$ oxygen-breathing phonon mode\cite{Lanzarakink} or  magnetic resonance mode\cite{Johnsonkink,Kaminskikink,KordyukT}, while a single $\sim$40 meV energy scale observed near the antinodal region is assigned to either magnetic resonance mode\cite{GromkoANKink} or B$_{1g}$ oxygen-buckling phonon mode\cite{CukANKink,DevereauxKink}. Besides the fact that the nature of the individual mode remains under debate, it has been difficult to understand how the nodal 70 meV kink can evolve into the anti-nodal 40 meV kink.  It is now clear that there are two separate energy scales that coexist over the large momentum space of the Brillouin zone. First, in the nodal region, the usually referred 70 meV kink is actually composed of two coexisting energy scales that are close in energy to each other. Second, the evolution of the nodal 70 meV to the antinodal 40 meV kink is not a single mode evolution, as previously thought. Instead, over the large momentum space, two energy scales coexist. The HI 78 meV energy scale does not change much in its energy while the LW 40 meV energy scale exhibit obvious energy shift.

Our present observations also pose a new challenge to our current understanding of mode coupling in the superconducting state in high temperature cuprate superconductors.  In a usual picture where the electron-boson coupling vertex g({\bf k},{\bf q}) is nearly momentum independent, if there is a mode (with an energy $\Omega$) coupling in the normal state, it is expected to be shifted to a new position $\Omega$+$\Delta_0$ in the superconducting state owing to the opening of the superconducting gap\cite{Sandvik}.  This isotropic coupling picture will generate an energy shift $\Delta_0$ over the entire Fermi surface, including the nodal region where the local superconducting gap is zero even in a {\it d}-wave superconductor like cuprates\cite{Sandvik}.  However, such a picture is not fully demonstrated in high temperature cuprate superconductors\cite{CukANKink,WSLee,Zhangfeatures}. As shown from the detailed temperature dependence of the electron self-energy (Fig. 3(g-i)), the HI mode already exists above T$_c$ at $\sim$78 meV. Upon entering the superconducting state, it stays nearly at the same energy and no signature of a new energy scale appears at 103 meV (with an energy gap of $\Delta_0$$\sim$25 meV for the OD82K Bi2212). The same is true for the Opt91K Bi2212 sample reported before\cite{Zhangfeatures}. While one may still attempt to explain the phenomena in a conventional picture\cite{WSLee}, it was proposed alternatively that this non-energy shift puzzle of the nodal 70 meV kink can be attributed to a strong momentum-dependent coupling vertex g({\bf k},{\bf q}) with a peculiar electron-boson coupling that involves mainly small {\bf $q$} forward scattering\cite{Kulic}. Recent observation of a low $\sim$10 meV energy scale along the nodal direction in the superconducting state of Bi2212\cite{Zhangfeatures,Vishikfeature,Plumbfeature} definitely cannot be explained by the usual energy-shift picture because its energy scale is smaller than the superconducting gap size ($\Delta_0$); in this case, the forward scattering picture provides a consistent explanation of this energy scale and its momentum dependence\cite{Johnston}. In the small-{\bf q} forward scattering picture, since electron scattering occurs in a small local momentum space,  a mode $\Omega$ in the normal state would be expected to be shifted by a local energy gap $\Delta$(k) upon entering the superconducting state\cite{Johnston}. A non-energy shift of the nodal kink is expected because the local gap near the nodal region is zero. It is also expected that the mode energy shift will increase from the nodal to the anti-nodal regions due to the local gap increase in a {\it d}-wave superconductor.  While the forward scattering picture\cite{Kulic,Johnston} seems to solve the non-shift puzzle of the nodal energy scale across T$_c$ that is difficult for the conventional isotropic coupling picture\cite{Sandvik}, it can not explain our new observation that the HI energy scale does not change in its energy over a large momentum space. The momentum dependence of the LW energy scale is even more anomalous because in this case, the energy scale change from 40 meV near the antinodal region to 70 meV near the nodal region is not only different from a constant that is expected from the conventional coupling picture, but is just opposite to an increase in energy from the nodal to the antinodal regions that is expected from the forward scattering picture. The present observations indicate that the electron-mode coupling in the superconducting state of Bi2212 is rather unusual that asks for further theoretical efforts.

In summary, by taking super-high resolution ARPES measurements, we have identified two coexisting sharp mode couplings  (HI scale at $\sim$78 meV and LW scale at 40$\sim$70 meV) over a large  momentum space in the superconducting state of  Bi2212 for the first time.  Although multiple mode coupling can be present in the normal state\cite{XJZhouFS,MeevasanaFS,LZhaoFS}, these two sharp modes  apparently stand out in the superconducting state due to their selectively much-enhanced coupling with electrons.  These two mode-couplings exhibit distinct momentum dependence; the HI mode keeps its energy near 78 meV over a large momentum space, while the LW mode changes its energy from $\sim$40 meV near the antinodal region to $\sim$70 meV near the nodal region, with its coupling strength dominant near the antinodal region. The observation of the coexisting two mode coupling provides a new insight on the mode coupling and their momentum evolution in Bi2212 superconductor. The unusual momentum dependence of these two mode-couplings can not be understood by known theories and poses new challenges to current understanding of electron-boson coupling in the superconducting state. Although our present work does not imply that the two modes are solely responsible for electron paring in the cuprate superconductors, it solves a long-standing puzzle about the evolution between nodal kink and antinodal kink, and their strong coupling in the superconducting state and their peculiar momentum and temperature dependence make it clear that such mode coupling should be considered in the electron pairing mechanism, be it pairing mediator or pairing breaker.

We thank Prof. Qianghua Wang for discussions. XJZ thanks the funding support from NSFC (Grant No. 11190022) and the MOST of China (Program No: 2011CB921703 and 2011CB605903).  The work at BNL is supported by the DOE under contract No. DE-AC02-98CH10886.

$^{*}$Corresponding author (XJZhou@aphy.iphy.ac.cn)


\newpage

\begin{figure*}[tbp]
\begin{center}
\includegraphics[width=0.92\columnwidth,angle=0]{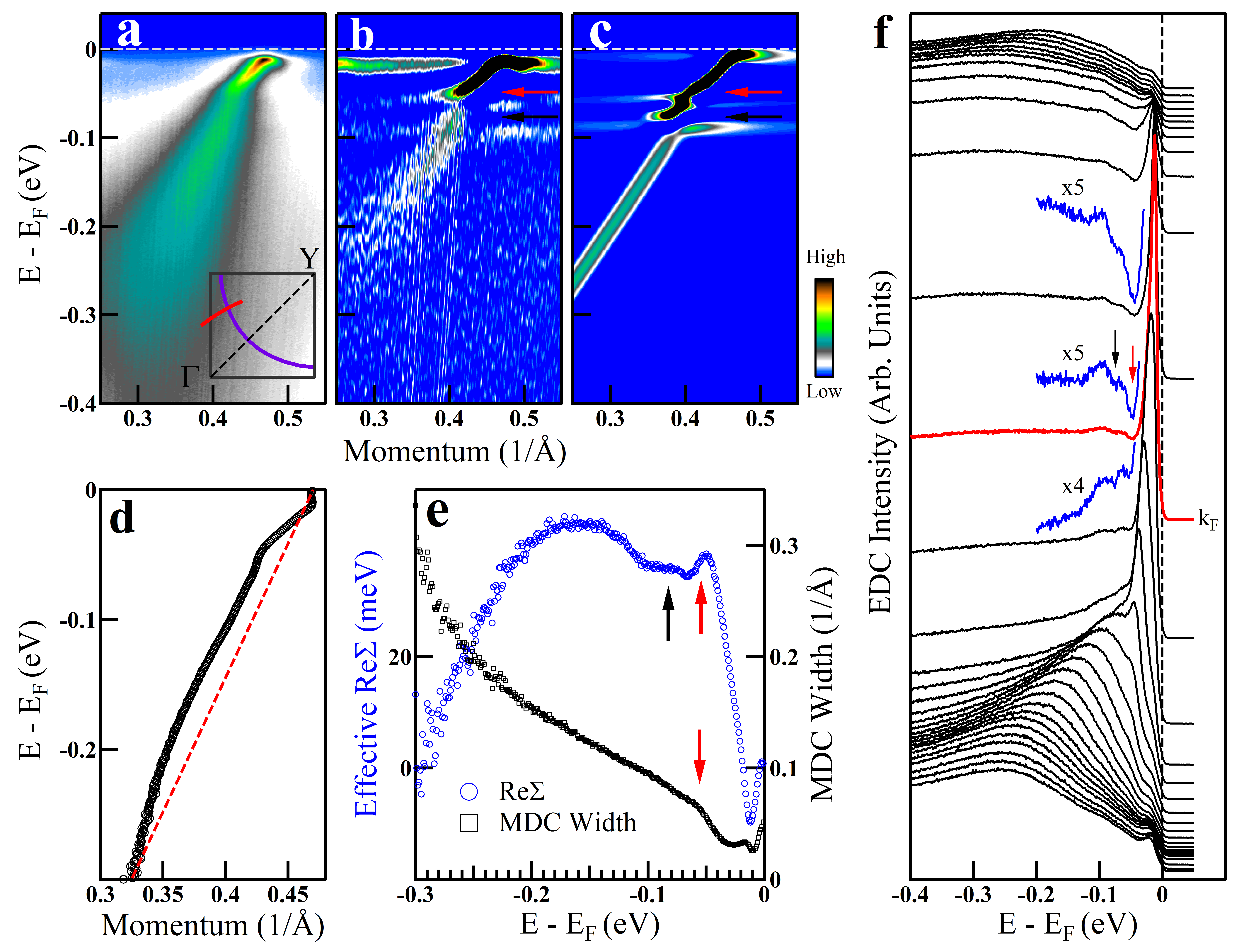}
\end{center}
\begin{center}
\caption{Identification of two coexisting energy scales in the Bi2212 OD82K sample.  (a). Original photoemission data measured at 17 K along a momentum cut marked in the inset.  (b). Corresponding second-derivative image with respect to energy in order to sharpen bands and fine features. Arrows point to the location of two energy scales. (c). Second-derivative image of the simulated single-particle spectral function which considers electron coupling with two Einstein phonon modes at 50 meV and 78 meV, as indicated by two arrows\cite{Supplementary}. (d). Energy dispersion extracted from fitting momentum-distribution curves (MDCs). (e). Effective real part of electron self-energy (blue circles), extracted from the measured dispersion in (d) by assuming a straight line connecting the -0.3 eV energy point on the dispersion and k$_F$ at zero energy as the empirical bare band (red dashed line in (d)),  and MDC width (full-width-at-half-maximum, FWHM, black squares). The features between the Fermi level and 30 meV are artifacts due to opening of superconducting gap. The arrows indicate positions of a sharp peak at $\sim$50 meV and another weak peak at $\sim$80 meV which are the focus of the present paper. (f). Photoemission spectra (EDCs) showing multiple peak-dip-hump structures. Typical  EDCs at the Fermi momentum (k$_F$) (red) and near k$_F$ are expended to show the structures more clearly and the arrows indicate the dip positions.}
\end{center}
\end{figure*}

\begin{figure*}[tbp]
\begin{center}
\includegraphics[width=1.0\columnwidth,angle=0]{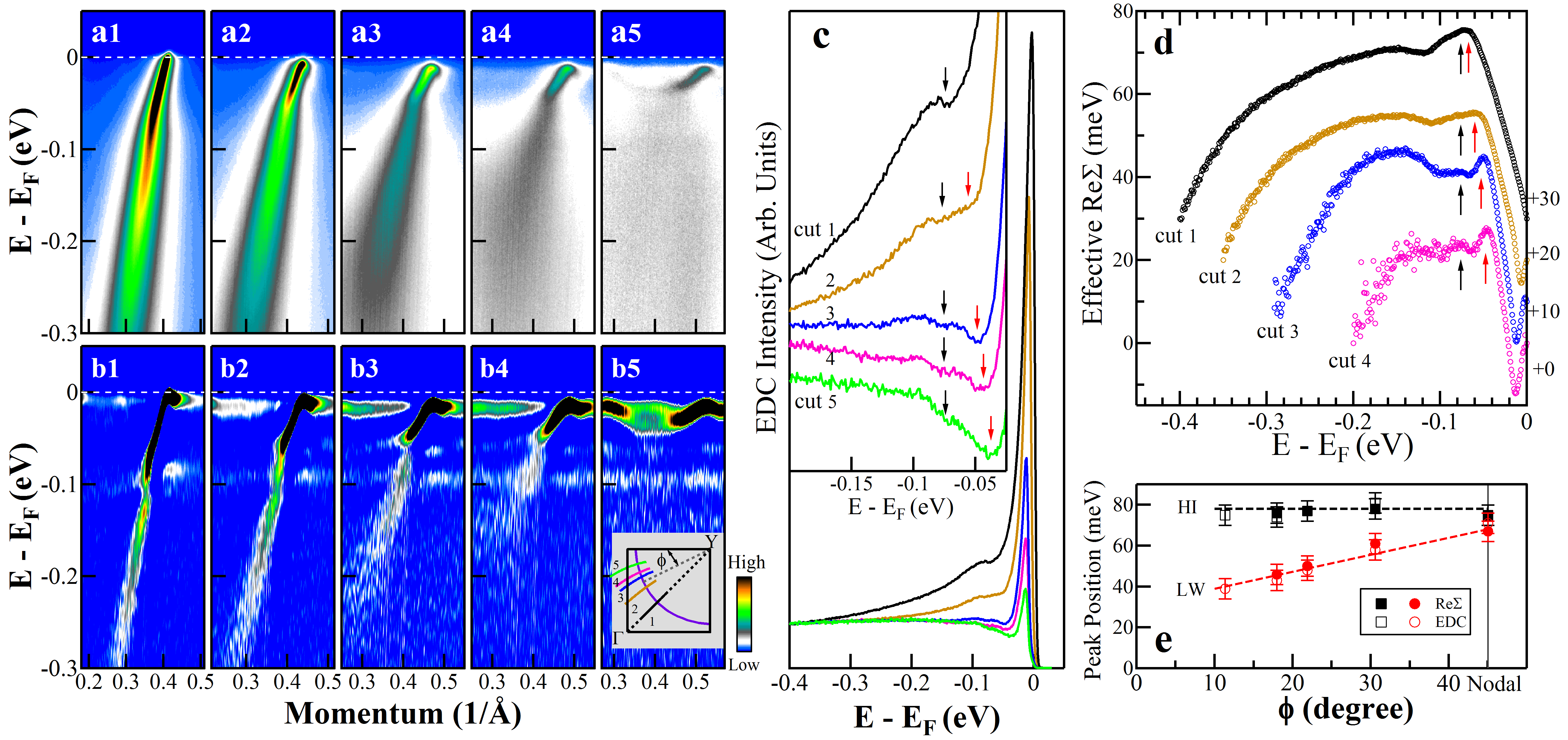}
\end{center}
\begin{center}
\caption{Momentum dependence of the two energy scales in Bi2212 OD82K sample. (a1-a5). Original images measured along five momentum cuts at 17 K. (b1-b5). Corresponding EDC second-derivative images. The location of the five cuts is shown in the inset. (c).  EDCs at the Fermi momenta k$_F$ of these five cuts. The EDCs in a smaller energy window are expanded in the top-left inset, and are offset vertically for clarity, to highlight the multiple peak-dip-hump structures.  Black and red arrows point to the two energy scales. (d). Effective real part of electron self-energy of the cut 1 to cut 4\cite{Supplementary}. For clarity, they are offset vertically with the offset values shown on the right side. (e). Momentum dependence of the two energy scales, extracted from the effective real part of electron self-energy (solid circles and squares) and dip position in EDCs (empty circles and squares). The $\Phi$ angle is defined in the inset of (b5) with $\Phi$=45 representing  the nodal direction.}
\end{center}
\end{figure*}

\begin{figure*}[tbp]
\begin{center}
\includegraphics[width=1.0\columnwidth,angle=0]{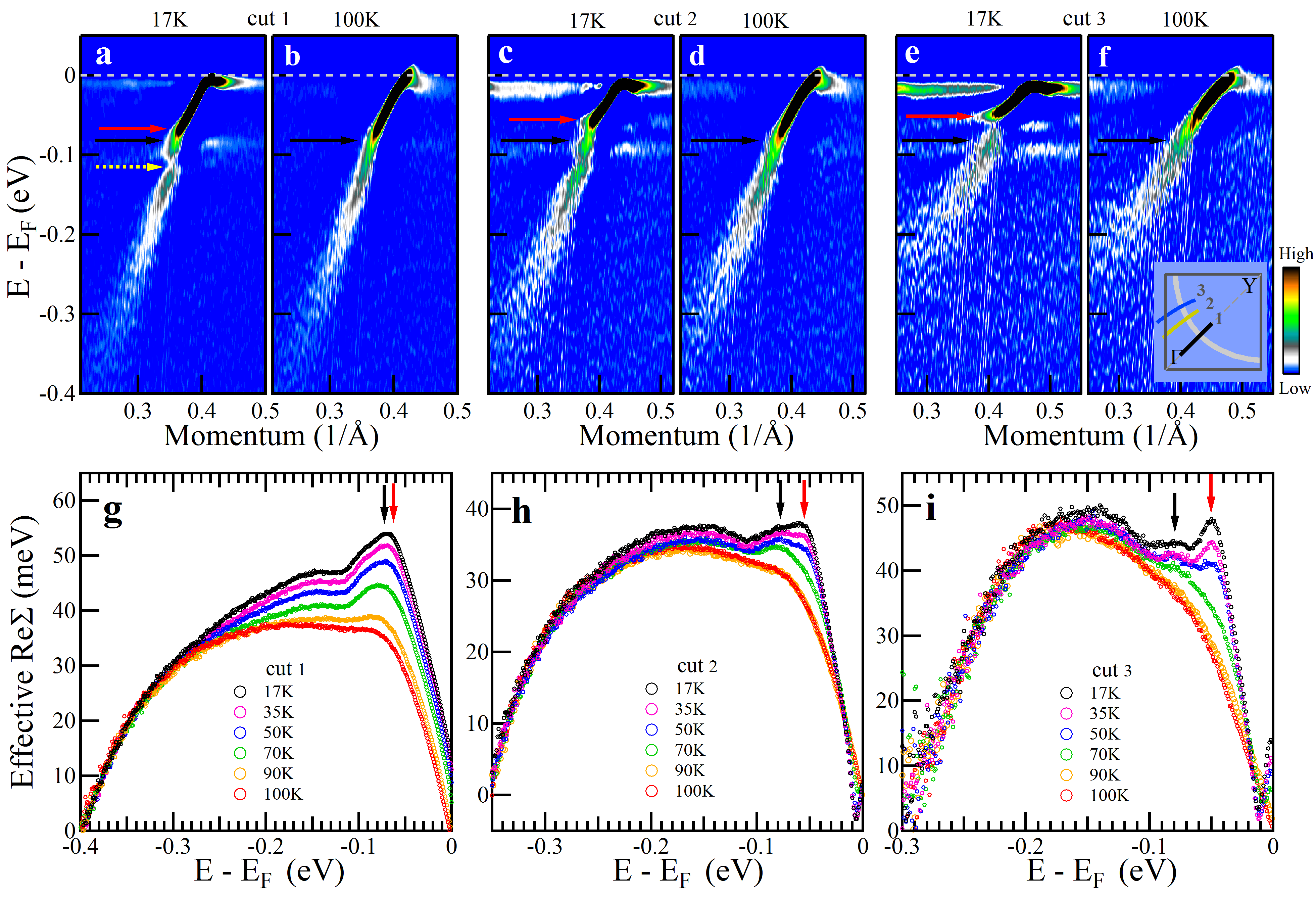}
\end{center}
\begin{center}
\caption{Temperature dependence of the two types of energy scales for the Bi2212 OD82K sample. (a,b) EDC second-derivative images of cut 1 measured below (at 17 K) and above (at 100K) T$_c$. (c,d) and (e,f) are the same measurements for cut 2 and cut 3 respectively. (g-i) Corresponding effective real part of electron self-energy, measured at different temperatures,  for the cut 1, 2, and 3, respectively.  For a given cut, a straight line connecting a high binding energy point on the 100 K dispersion and its k$_F$ at the Fermi level is taken as a common empirical bare band for all the temperatures\cite{Supplementary}. Peaks of two energy scales are marked by arrows, with the red one representing the LW energy scale while the black one representing the HI energy scale.}
\end{center}
\end{figure*}

\begin{figure*}[tbp]
\begin{center}
\includegraphics[width=0.92\columnwidth,angle=0]{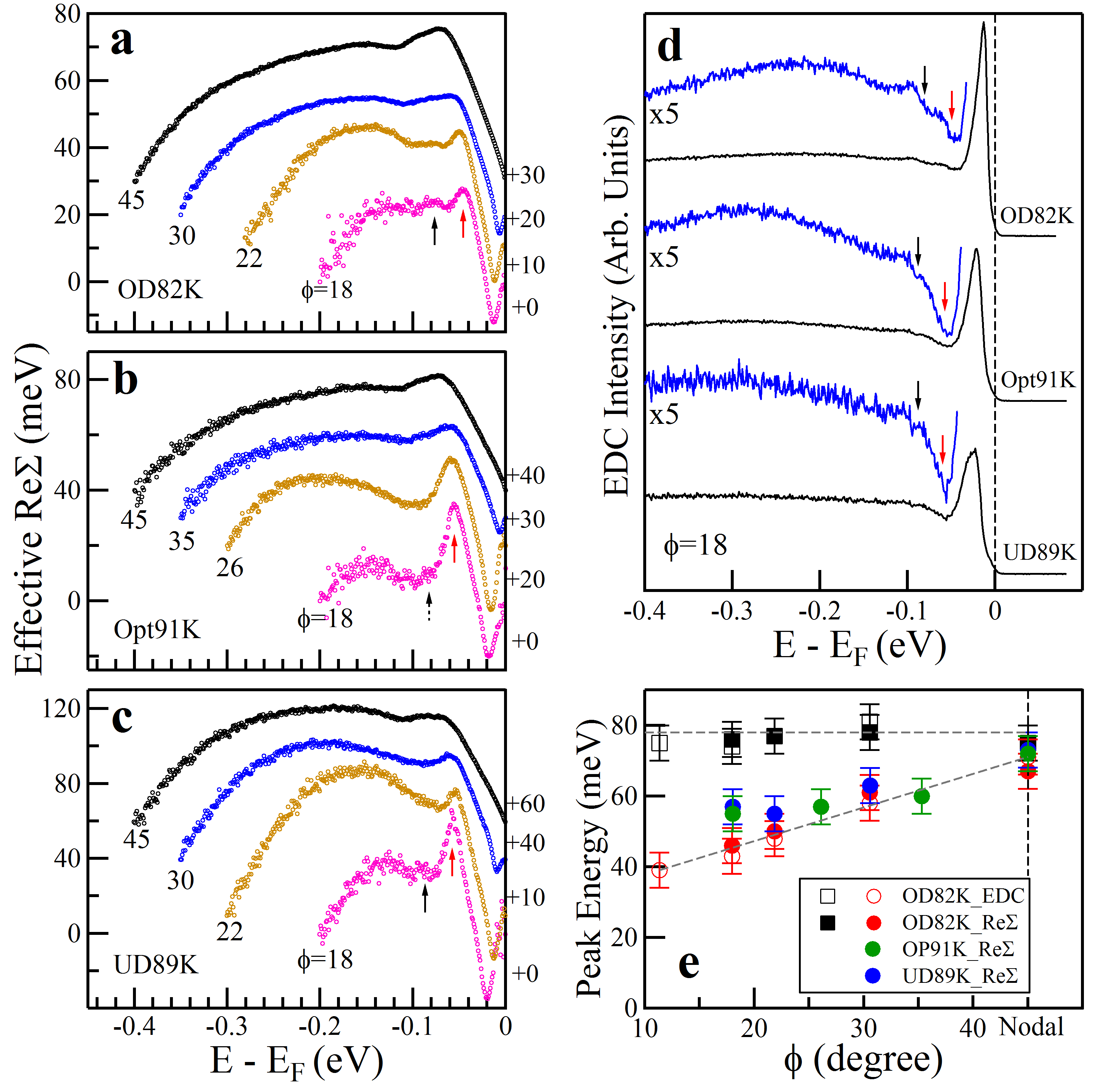}
\end{center}
\begin{center}
\caption{Doping dependence of the two types of energy scales in Bi2212 samples.  Effective real part of electron self-energy measured in superconducting state for OD82K sample (a), Opt91K sample(b) and UD89 sample(c)\cite{Supplementary}. The location of the momentum cuts are given by $\Phi$ angles as defined in Fig. 2b5. For clarity, the effective real part of electron self-energy is offset vertically with the offset values marked on the right side of the curves. (d) EDCs at the Fermi momentum k$_F$ along the cut $\Phi$=18 degree for the three samples with different doping levels. Part of the EDCs is expanded (blue curves) in order to highlight the possible multiple peak-dip-hump structures. Arrows mark the position of dips. (e). Summary of the momentum dependence of the two types of energy scales in Bi2212 with different doping levels. Both EDC and electron self-energy are used to obtain the energy positions.}
\end{center}
\end{figure*}

\end{document}